\documentclass{article}
\usepackage{graphicx} 
\usepackage[a4paper, total={6in, 8in}]{geometry}

\usepackage{booktabs}
\usepackage{multicol}
\usepackage{fancyhdr}
\usepackage[hidelinks,colorlinks=true,linkcolor=blue,citecolor=blue]{hyperref}
\usepackage{amsmath}	
\usepackage{amssymb}	
\usepackage[inter-unit-product=\cdot]{siunitx}
\usepackage{enumerate}
\usepackage{multirow}
\usepackage{mathtools}
\usepackage{comment}
\usepackage{longtable}
\usepackage{tabularx}
\usepackage{xcolor} 
\usepackage{ulem} 
\usepackage{comment}
\usepackage{cuted}
\usepackage{soul}
\usepackage{authblk}

\newcommand{\Msun}{\,\mathrm{M}_\odot}
\newcommand{\Rsun}{\,\mathrm{R}_\odot}

\newcommand{\tonde}[1]{\left(#1\right)}
\newcommand{\quadre}[1]{\left[#1\right]}
\newcommand{\graffe}[1]{\left\{#1\right\}}

\DeclareSIUnit\year{yr}
\DeclareSIUnit\au{AU}
\DeclareSIUnit\parsec{pc}
\DeclareSIUnit\erg{erg}

\newcommand{\orcidicon}[1]{\href{https://orcid.org/#1}{\includegraphics[width=11pt]{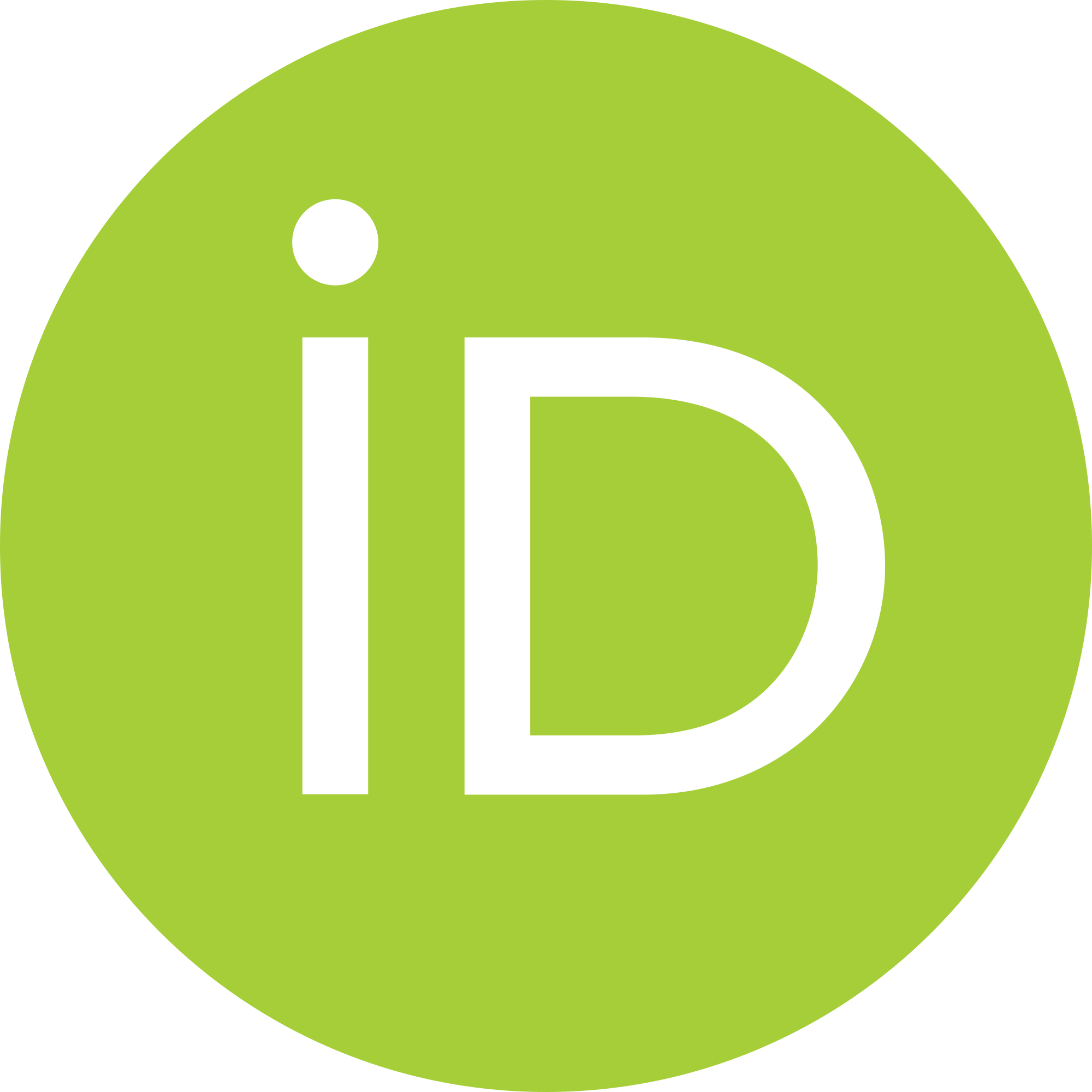}}}
\newcommand{\orcid}[1]{\href{https://orcid.org/#1}{\protect\orcidicon{#1}}}

\title{Astrophysics: a Modern Discipline with a Newtonian origin
}
\author{Maria Paola Vaccaro \orcid{0000-0003-3776-9246} }
\affil{Institut f{\"u}r Theoretische Astrophysik, ZAH, Universit{\"a}t Heidelberg, Albert-Ueberle-Stra{\ss}e 2, D-69120, Heidelberg, Germany\\  \href{mailto:mariapaolavaccaro@gmail.com}{mariapaolavaccaro@gmail.com}}
\date{March 2025}

\begin{document}

\maketitle

\begin{abstract}
Understanding the formation and evolution of stellar-mass binary black holes (BBHs) requires a thorough investigation of the key physical processes involved. While one pathway involves the isolated evolution of massive binary stars, affected by uncertain stages like core-collapse supernovae and common envelope evolution, an alternative channel is dynamical formation in dense stellar environments. Newtonian gravity has traditionally provided a robust and computationally efficient framework for modeling large-scale gravitational interactions. However, accurately capturing close encounters and black hole mergers necessitates the use of general relativity. This work focuses on assessing the applicability of post-Newtonian gravity in bridging these regimes, offering a physically insightful and computationally tractable approach to modeling BBH formation in the gravitational-wave era of astronomy. \\
\\
\noindent \textbf{This is the author’s accepted manuscript} of the article:  
M. P. Vaccaro,``Astrophysics: a modern discipline with a Newtonian origin", \textit{Phil. Trans. R. Soc. A}, 383:20230294 (2025).  
This article is part of the theme issue ‘Newton, Principia, Newton Geneva Edition (17th–19th) and modern Newtonian mechanics: heritage, past \& present’.\\
© The Author(s) 2025. Published by the Royal Society. All rights reserved.

\end{abstract}

\section{Introduction}
Astronomy, as one of the oldest sciences, has its theoretical foundation firmly grounded in Newtonian gravity. Developed by Isaac Newton in the 17th century, this theory of gravitation successfully explained a wide range of astronomical phenomena, from the orbits of planets around the Sun to the motion of stars within galaxies \cite{Principia}. Newtonian gravity remains a powerful tool for understanding the large-scale dynamics of the universe, where the effects of gravity are significant but not extreme.

In the last decade, the LIGO and Virgo interferometers have directly detected gravitational waves from mergers of stellar-mass black hole pairs, as well as a smaller number of mergers involving neutron stars. These discoveries offer the exciting possibility that compact object mergers could serve as valuable probes of stellar and binary evolution, and potentially even of stellar dynamics.

To understand the formation of  stellar-mass   binary black holes (BBHs) and their eventual mergers, it is crucial to review the main physical processes involved. One formation pathway is through the isolated evolution of massive binary stars. However,  some of the stellar evolution processes are still uncertain  , particularly regarding the physics of core-collapse supernovae and the common envelope phase, both of which significantly influence the final system.
An alternative route is the dynamical formation channel, where two black holes come together through gravitational interactions in dense star clusters. Newtonian gravity is particularly well-suited to modeling these large-scale, complex interactions in dynamic environments, as it provides a robust framework for understanding the movement and encounters of objects within such clusters. However, the intense gravitational fields that occur during close encounters require the more accurate framework of general relativity (GR) \cite{GR_1915}  . 

While GR governs the physics of gravitational waves and black hole mergers, Newtonian dynamics remains a crucial tool for modeling the environments in which these events unfold. This paper explores how Newtonian methods, despite their simplicity, provide computationally efficient and physically insightful approaches to understanding black hole formation in dense stellar environments. This is crucial to understand how different   formation pathways leave distinct imprints on the mass, spin, and orbital properties of the resulting BBHs, providing unique clues about their origins.

\section{Gravitational wave observations}
Recent observations of gravitational waves have provided invaluable information on the population of black holes in the Universe. 
In this section we will briefly introduce the nature of black holes, and discuss gravitational waves emission and detection.

\subsection{Black Holes}
\label{sec: intro BH}
A black hole (BH) is an incredibly compact 
object such that all of its mass $M_\mathrm{BH}$ is contained inside its event horizon. For a non-spinning BH, the radius of the event horizon is defined as the Schwarzshild radius 
$R_S= 2GM_\mathrm{BH}/c^2$, where $G$ is the gravity constant, $c$ the speed of light and $M_{\rm BH}$ is the mass of the BH. 
At the event horizon, the gravity is so strong that nothing, not even photons, can escape from it. 


Few people know that the idea of the existence of BHs was first postulated in 1783 by Rev. John Michell \cite{John_Michell}. He argued that a sufficiently compact star may have a surface escape velocity larger than the speed of light and would thus be invisible. He called these objects `dark stars'. This idea had very little traction on his contemporaries and was forgotten until his writings re-surfaced in the 1970s \cite{article_Mitchell}. 

A proper mathematical treatment of BHs had to await 1916, when Albert Einstein formulated the theory of GR and, shortly after, Karl Schwarzschild
found a spherically symmetric solution of Einstein's equations in the vacuum. He demonstrated the existence of a characteristic event horizon, the Schwarzschild radius $R_S$, within which no communication is possible with external observers. Later, Roy Kerr generalized this solution to spinning black holes in 1963 and finally John Wheeler coined the term `black hole' in 1968. \cite{Genzel_2021_SMBH}

Nowadays, we have astronomical evidence for the existence of tens of BHs in the local Universe, and we classify them based on their mass: stellar BHs with masses lower than $100\Msun$, 
    intermediate BHs with masses between $100\Msun$ and $10^5\, \Msun$,
    and supermassive BHs with masses larger than $10^5\,\Msun$,

 where $\Msun$ is the mass of our Sun, roughly equal to $2\times 10^{30}$ kg.  

Stellar BHs can form directly from the collapse of massive stars (with initial mass $M_*\gtrsim 20 \Msun$). The theory of stellar evolution predicts that a phase of hydrogen burning, which occupies most of a star's life and which is currently ongoing in our Sun, is followed by a period in which the outer layers of the star expand while its core continues to contract and heat, enabling the fusion of heavier elements (helium, carbon, oxygen, and so on, up to iron). Once the core is primarily composed of iron, nuclear fusions can no longer produce the energy necessary to sustain the structure of the core, so that it collapses under its own gravity into a BH, while the outer layers of the star are expelled in a supernova explosion \cite{Woosley_2002}.

Moreover, we expect a gap in BH masses between $\sim 60 \Msun$ and $120 \Msun$  \cite{Spera_2017}  , which is referred to as the upper mass gap or pair-instability mass gap. The reason for this gap is that very massive stars (with helium core masses between $64 \Msun$ and $133 \Msun$ \cite{PISN}) experience an instability in their core due to the creation of electron-positron pairs out of two photons. This leads to a runaway mechanism which is predicted to produce a very powerful explosion, called pair-instability supernova, that disrupts the entire star and leaves no compact remnant.
The lower limit of the mass gap has large uncertainty due to our poor understanding of the physics of massive stars and, in particular, of the ${}^{12} \mathrm{C} \tonde{\alpha, \gamma} {}^{16}\mathrm{O}$ reaction rate \cite{Costa_et_al_2020, Farmer_2019}.

\subsection{Gravitational Waves}

According to the theory of GR, mass is the source of space-time curvature. A massive object in an accelerated motion produces ripples in spacetime, called gravitational waves (GWs), which propagate away at the speed of light \cite{Hartle_2013}.

In this section we will discuss the emission of gravitational waves by binary systems, their detection and some astrophysical implications.

\subsubsection{GW emission 101}

We consider a binary system made of two compact objects, such as neutron stars or black holes, of masses $m_1$ and $m_2$  , treating them   as point-like masses. Assuming circular orbits  (i.e. with null eccentricity, $e=0$)  , their dynamics is equivalent to that of a body of mass $\mu=m_1 m_2/\tonde{m_1 + m_2}$ on a circular orbit of radius $R$ and orbital frequency $\omega_s^2 = G \tonde{m_1 + m_2}/R^3$.

It is useful to define the chirp mass $M_c$ as
\begin{equation}
    M_c =\frac{\tonde{m_1m_2}^{3/5}}{\tonde{m_1+m_2}^{1/5}}
    \label{eq:chirp_mass}
\end{equation}

 This is a key parameter in GW astronomy, encapsulating a binary's mass combination in a way that directly influences the frequency evolution of the emitted waves. It also governs the amplitude and rate of frequency change of the signal, making it critical for waveform modeling.  

The GW strain\footnote{The strain is $h=\delta L/L$, where $L$ is the distance between two reference points in space and $\delta L$ is the induced displacement between them. \cite{roadmapGW_Bailes}} at lowest order (quadrupole) generated from such a binary, for the two polarizations $+$ and $\times$, is \cite{maggiore1_GW}
\begin{align}
    h_+\tonde{t} &= A\tonde{M_c, \omega_s, r}\, \frac{1+\cos^2{\theta}}{2} \cos{\tonde{2 \omega_s t_\mathrm{ret} +2\phi}} 
    \label{eq:h_+}\\
    h_\times \tonde{t} &= A\tonde{M_c, \omega_s, r}\, \cos{\theta} \sin{\tonde{2 \omega_s t_\mathrm{ret} +2\phi}} 
    \label{eq:h_x}
\end{align}
where 
\begin{equation}
    A\tonde{M_c, \omega_s, r} = \frac{4}{r} \tonde{\frac{G M_c}{c^2}}^{5/3} \tonde{\frac{\omega_s}{c}}^{2/3} ,
\end{equation}
$r$ is the distance from the observer, $t_\mathrm{ret}=t-r/c$ is the retarded time, and $\theta$ and $\phi$ are the zenith and azimuth angle at which the observer is viewing the source. We notice that the GW frequency is equal to twice the orbital frequency of the source: $\omega_\mathrm{gw} = 2\, \omega_s$.

GWs carry energy away from the source. The total radiated power is approximately \cite{maggiore1_GW}
\begin{equation}
    P_\mathrm{gw} = \frac{32}{5} \frac{c^5}{G} \tonde{\frac{G M_c \omega_\mathrm{gw}}{2c^3}}^{10/3} 
    \label{eq:power_radiated_gw}
\end{equation}
We assume that the point masses have no internal structure, so the only possible source of energy is the orbital energy of the binary $E_\mathrm{orbit}$. Even in a realistic system of two extended objects, corrections due to their internal structure are of order $\tonde{v/c}^{10}$ and can safely be neglected for non-relativistic motion \cite{maggiore1_GW}.

The orbital energy and its time evolution are 
\begin{equation}
    E_\mathrm{orbit} = - \frac{G m_1 m_2}{2R}\,, \qquad\qquad \frac{d E_\mathrm{orbit}}{dt} = -P_\mathrm{gw}
    \label{eq:GW_energy_loss}
\end{equation}

So, if $E_\mathrm{orbit}$ decreases to compensate for the loss of energy to GWs, $R$ must decrease as well. If $R$ decreases, $\omega_s$ increases and, in turn, the GW strain $h_{+,\times}$ increases. This leads to a runaway process where the binary keeps shrinking in time and the amplitude of GWs keeps increasing until, eventually, the binary reaches coalescence. 


In truth, the expressions for the GW strain in eq.s \ref{eq:h_+} and \ref{eq:h_x} and for the radiated power in eq. \ref{eq:power_radiated_gw} are only valid in the approximation of quasi-circular orbits ($\dot\omega_s \ll \omega_s^2$ , $e=0$  ). When a binary shrinks, it eventually reaches a condition where this approximation is no longer valid. In particular there is a minimum value of the radial distance beyond which stable circular orbits are no longer allowed. This is called the Innermost Stable Circular Orbit (ISCO) and, for non rotating BHs and $m_1 \gg m_2$, it is equal to 
\begin{equation}
    R_\mathrm{ISCO}= \frac{6\, G \tonde{m_1+m_2}}{c^2} = 3\, R_S,
    \label{eq:GW_R_ISCO}
\end{equation}
where $R_S = 2GM/c^2$ is called Schwarzschild radius.

Therefore, for $R\gtrsim R_\mathrm{ISCO}$, the binary goes through a succession of quasi-circular orbits with progressively decreasing radius. Then, when the orbital separation approaches $R_\mathrm{ISCO}$,  the system enters   a plunge phase where the radius decreases quickly and a correct estimate of the  orbital   evolution requires numerical relativity.

 The quasi-circular orbits approximation also fails if the binary has a non-zero initial eccentricity, as the orbital motion becomes more complex and higher harmonics contribute to the gravitational-wave signal.  



\subsubsection{GW detections and astrophysical implications}
\label{sec:GW_detection}

Recent observations of gravitational waves have provided invaluable information on the population of BHs in the Universe. This has been possible thanks to the construction of four GW detectors: two LIGO detectors in the USA, the Virgo detector in Italy and the KAGRA detector in Japan. They are advanced modified Michelson interferometers with arms of length of 3 to 4 $\si{\kilo\meter}$. They are sensitive in the frequency range from $\sim \SI{10}{\hertz}$ up to a few $\si{\kilo \hertz}$. We will refer to the LIGO--Virgo--KAGRA Collaboration as LVK.

The LVK interferometers are incredibly sensitive to spatial distortions: typical strains from astrophysical sources are on the order of $\delta L/L = 10^{-21}$ or less \cite{roadmapGW_Bailes}. Thus, with arms with length of the order of a few kilometers, the interferometers must be sensitive to displacements $\delta L$ of less than $\sim 10^{-18}\, \si{\meter}$. This is an incredibly small displacement: for comparative purposes, note that the radius of a proton
is $\sim \SI{8.4e-16}{\meter}$ \cite{proton_radius_2022}. This high sensitivity is achieved by using Fabry-Perot resonant cavities, so that the optical length of the arms is of a few hundred kilometers.


 The emitted frequency and amplitude of the GW signal depends on the properties of the BBH, for example on the primar and secondary masses $m_1$ and $m_2$. More massive binaries produce lower maximum frequencies and shorter signals in the LVK detectors, not because their inspirals are brief, but because much of the inspiral occurs at frequencies below the detectors' sensitivity range.   

For example, the signal of a binary with masses $m_1=m_2=1.4\Msun$ enters the detector's bandwidth ($f_\mathrm{gw} \geq \SI{10}{\hertz}$) at $\tau=\SI{17}{\minute}$ prior to coalescence and reaches a maximum frequency $f_\mathrm{max} = \SI{1.6}{\kilo\hertz}$. If we increase the masses up to $m_1=m_2=10\Msun$, the signal is visible for only $\tau=\SI{38}{\second}$ and the maximum frequency is $f_\mathrm{max} = \SI{440}{\hertz}$, which is well within the detector's bandwidth. For an even more massive binary with $m_1=m_2=100\Msun$, the signal is visible for only $\tau=\SI{8}{\second}$ and it reaches a maximum frequency of $f_\mathrm{max}=\SI{22}{\hertz}$, which is in the higher-noise region of the bandwidth 
Since it is not possible to identify signals that are too short or that have a low signal-to-noise ratio, the LVK interferometers are not sensitive to mergers of high-mass compact binaries.

The most used technique to identify astrophysical GW signals is `matched filtering', which uses a bank of strain templates computed using numerical relativity and searches for similar patterns in the data \cite{GWTC3_first}. Each template encodes the properties of the merging BBHs: the component masses and spins. Coupling
information from two or more detectors in a network, it is also possible to infer the sky location and orientation of
the source, as well as our distance to the source \cite{MANDEL20221}.

The templates are well characterized by combinations of the binary component parameters such as the chirp mass $M_c$ (eq. \ref{eq:chirp_mass}), the mass ratio $q=m_2/m_1$ and two combinations of the BH spins called the effective spin $\chi_\mathrm{eff}$ and the precession spin $\chi_\mathrm{p}$. These quantities determine the phase evolution during inspiral \cite{GWTC-1}. 

The expression for the effective spin is
\begin{equation}
    \chi_\mathrm{eff} = \frac{m_1 \vec\chi_1 + m_2 \vec\chi_2}{m_1+m_2}\cdot \frac{\vec L}{\lvert \vec L \rvert}
    \label{eq: effective spin}
\end{equation}
where $\vec L$ is the orbital angular momentum and $\vec\chi_i$ ($i=\graffe{1,2}$) are the dimensionless spins, defined as $\vec\chi_i = c \vec S_i / G m_i^2$ where $\vec S_i$ is the spin angular momentum. For BHs, $\chi_i$ can theoretically range from $0$ (non-spinning) to $1$ (maximally-spinning). 

The expression for the precession spin is
\begin{equation}
    \chi_\mathrm{p} = \max\graffe{\chi_{1,\perp},\, \frac{q\tonde{4q+3}}{4+3q} \chi_{2,\perp}} 
    \label{eq: precession spin}
\end{equation}
where $\chi_{i,\perp}$ is the component of spin perpendicular to the direction of the orbital angular momentum $\vec L$ ($i=\graffe{1,2}$).

Hence $\chi_\mathrm{eff}$ is a measure of the BH spin components along the orbital angular momentum vector, while $\chi_\mathrm{p}$ measures the spin components in the orbital plane. The precession spin owes its name to the fact that the presence of spin components perpendicular to the orbital angular momentum vector causes a precession of the binary orbital plane.



From the beginning of observations in September 2015 up until the end of the third observation run (O3b) in March 2020, the LVK collaboration has detected \cite{GWTC3_second} 76 transient signals associated with the inspiral and merger of compact binaries with false alarm rate $\leq \SI{1}{\year^{-1}}$. Of these, 13 merger events have at least one of the progenitors with mass overlapping with the $\quadre{\sim 60, 120}\Msun$ range, usually referred to as the \textit{pair-instability} or \textit{upper} mass gap, at 90\% credible interval. 
This poses a direct challenge to standard stellar evolution models, suggesting either that current models of pair-instability supernovae need revision — allowing for the formation of BHs in the upper mass gap — or that these BHs originate through dynamical processes such as hierarchical mergers in dense environments.

 Although the final inspiral and merger of BBHs are inherently relativistic events, their dynamical history — from formation to hardening — often unfolds in environments where Newtonian gravity provides an adequate description. This justifies the use of $N$-body simulations based on post-Newtonian corrections to model BH dynamics in stellar clusters.  

\section{Newtonian aspects in an intrinsically post-Newtonian\\ phenomenon}
\label{sec:Nbody}

 In an effort to constrain the origin of observed BBH mergers, numerous studies \cite{Antonini_2020, Antonini_2023, Antonini_2016, Arcasedda_2023, Kritos_2023, Fragione_2023, fastcluster2021, fastcluster2022, Vaccaro_2023} have focused on modeling stellar dynamics in complex environments such as stellar clusters. In this section we will explore how, although GWs are inseparable from the the concept of spacetime, we can generally employ Newtonian gravity. We will examine why this is a reasonable approach in many contexts.

\subsection{Modeling gravitationally-bound systems: a Newtonian approach}

In order to model the behavior of an $N$-body system such as a star cluster, astrophysicists need to solve a set of equations of motion for $N$ bodies subject to Newton's gravity force, namely
\begin{equation}
    \frac{\mathrm{d}^2 \vec x_i}{\mathrm{d}t^2} = -G \sum_{j=1, j\neq i}^{j=N} m_j \frac{\vec x_i - \vec x_j}{|\vec x_i - \vec x_j|^3}
    \label{eq:Newton}
\end{equation}
where $\vec x_i$ is the position vector of the $i$th particle in the
 system, $m_j$ is the mass of the $j$th particle and $G \simeq 6.667 \times 10^{-8}$ cm$^3$ g$^{-1}$ s$^{-2}$ is the gravity constant. In this scenario, each particle is understood to be a star or a star remnant, and their internal structure is neglected.

The above equation can be split into a system of two first-order differential equations:
\begin{equation}
    \frac{\mathrm{d} \vec v_i}{\mathrm{d}t} = -G \sum_{j=1, j\neq i}^{j=N} m_j \frac{\vec x_i - \vec x_j}{|\vec x_i - \vec x_j|^3} ,
\end{equation}
\begin{equation}
    \frac{\mathrm{d} \vec x_i}{\mathrm{d}t} = \vec v_i
\end{equation}
where $\vec v_i$ is the velocity of the $i$th particle.

These are vector equations, which means they must be solved separately for the three spatial component and for every particle of our system. Hence, integrating the equation of motion of the entire $N$-body system requires the simultaneous solution of $6\times N$ first-order differential equations.  

This is performed numerically by means of numerical integrators, such as the Euler's method. In this scheme, one assumes the initial positions $\vec x_i$ and velocities $\vec v_i$ at a certain time $t$, and then evaluate them after a certain time interval $h$ as
\begin{equation}
    \vec a_i (t) = -G \sum_{j=1, j\neq i}^{j=N} m_j \frac{\vec x_i - \vec x_j}{|\vec x_i - \vec x_j|^3} ,
\end{equation}
\begin{equation}
    \vec x_i (t+h) = \vec x_i (t) + \vec v_i(t) h +\mathcal{O}(h^2),
\end{equation}
\begin{equation}
    \vec v_i (t+h) = \vec v_i (t) + \vec a_i(t) h +\mathcal{O}(h^2),
\end{equation}
where $\vec a_i = \mathrm{d} \vec v_i / \mathrm{d}t$ is the acceleration of the $i$th particle.

Repeating this computation for $M$ time steps, will provide the dynamics of the system during a time interval $T=Mh$. This is not an exact computation: the errors scale as $h^2$. The results can be improved by reducing $h$, but if $h$ is too small the calculation can become extremely slow.

In the case of a star cluster, where we want to model the dynamics of millions of stars, astrophysicists need to implement more sophisticated numerical techniques than directly solving the equations listed above. The most commonly-used direct $N$-body code is called \textsc{NBODY6++} and it requires a supercomputer \cite{Wang_2015}.

\subsection{ To GR or not to GR? The case for Newtonian approximations  }

The GR equation that should be solved  for a proper treatment of gravitational interactions   reads
\begin{equation}
    G_{\mu \nu} = \frac{8\pi G}{c^4} T_{\mu \nu}  ,
\end{equation}
which dictates how matter (described by the stress–energy  tensor $T_{\mu \nu}$) determines the curvature of spacetime ( represented by the Einstein tensor  $G_{\mu \nu}$). Here, $G \simeq 6.667 \times 10^{-8}$ cm$^3$ g$^{-1}$ s$^{-2}$ is the gravity constant and $c=2.998 \times 10^{10}$ cm s$^{-1}$ is the speed of light.

This equation is quite difficult to solve exactly and analytically, with the exception of a few special cases involving particular geometric symmetries.
More often, scientists  resort to perturbative methods to get the approximate solutions. One of the most widely used approaches is the post-Newtonian expansion , which is applicable in weak-field and slow-motion regimes \cite{Blanchet_2014}.

In the post-Newtonian scheme, we can expand relativistic tensors in small parameters,  such as the ratio of the velocity of
 the matter to the speed of light, $v/c$, or the ratio between the distance from the object and its Schwarzschild radius, $r/ R_S$\footnote{ The Schwarzschild radius $R_S$
  is the radius of the event horizon of a non-rotating black hole, and it is given by $R_S=2GM/c^2$ for a BH of mass $M$.} These expansions capture deviations from classical Newton's law. Thus, the zeroth-order expansion gives the classical Newton's law ,while higher-order Post-Newtonian  terms account for relativistic corrections and
 are valid in the case of weak fields (at a large distance from a BH or other gravitational source) and slow motion. 

To show this explicitely, let us put ourselves in a scenario where we have a single, non-rotating point mass $M$. In GR, this is called a Schwarzschild metric and the gravitational potential at a distance $r$ from the point mass is expressed as \cite{Rindler_2010}
\begin{equation}
\Phi = \frac{c^2}{2} \log\left( 1- \frac{2GM}{rc^2} \right)
\end{equation}

At large distances from the mass, $2GM/rc^2 = R_s/r$ is a small quantity , so the logarithm can be approximated  using the series expansion for $\log\tonde{1-x}$, valid when $x\ll1$:
\begin{equation}
    \log\left( 1- x \right) = x -\frac{x^2}{2} +\mathcal{O}\left(x \right)^3 .
\end{equation}
Hence, the gravitational potential can be expressed as
\begin{equation}
    \Phi = -\frac{GM}{r} -\frac{1}{c^2}\left(\frac{Gm}{r} \right)^2 +\mathcal{O}\left(\frac{GM}{r} \right)^3 .
\end{equation}

It is  clear that the leading order in the weak fields approximation, $\Phi = -GM/r $ is the classical Newtonian gravitational potential. This allows to neglect relativistic effects between two particles in a stellar clusters, unless they experience a close encounter. 

 However, higher-order post-Newtonian terms become crucial for accurate modeling of close gravitational systems. For instance, in the case of BBH systems, terms at least up to 3.5 post-Newtonian order are needed to accurately model the gravitational waveform as the system evolves toward merger \cite{Trani_2023}. These higher-order corrections account for phenomena such as energy loss due to gravitational radiation and more subtle relativistic effects like spin-orbit coupling and spin-spin interactions.

\section{ Unveiling the origins of binary black holes}
BBH systems can form through two primary channels: original or dynamically assembled binaries.

An original BBH comes from the evolution of a binary of two massive stars, where the stars undergo core collapse and create BH remnants,  assuming that the binary is not disrupted by the supernova explosions,  and its modeling requires detailed prescriptions for stellar evolution. Instead,  dynamically assembled BBHs form in dense stellar environments, such as stellar clusters,  which can be modeled by means of $N$-body simulations (see \autoref{sec:Nbody}).

\subsection{ The issues with isolated evolution}
The detection of GWs from BBH mergers proves
the existence of BBHs with a very short orbital separation, since the
initial separation of a BBH must be of the order of few ten
solar radii for the BBH to merge within a Hubble time\footnote{The Hubble time is defined as the time required for the Universe to expand to its present size, assuming that the rate of expansion has remained unchanged since the Big Bang, and it roughly corresponds to the age of the Universe. It is defined as the reciprocal of the Hubble constant, $1/H_0$, and it is around 14 billion years according to current measurements of $H_0$.} by
GW emission.  This presents a challenge for binary star evolution, as the parent stars of the BHs cannot be so close to each other. Indeed, stars expand during their evolution: the Sun will inflate up to a size of roughly $200 \Rsun$ after hydrogen is exhausted from its core, and stars which produce BHs as their remnants (of mass greater that $20 \Msun$) may reach an extension of roughly $1000 \Rsun$ \cite{Hurley_2000}. 

There seems to be an issue. If the the stellar binary is tight enough for GWs to eventually bring their remnants together, the stars would expand to sizes much larger than their separation during their evolution, likely causing them to merge long before producing BHs. If, instead, the stellar binary is loose enough to avoid merging before the production of BHs, their remnant binary system would take vastly longer than the Hubble time to merge. Either way, no GW sources would be detectable today.

 To reconcile this, one must account for the processes that can bring stars closer as they expand and interact.  Key processes 
include stable mass transfer via Roche lobe overflow (RLOF) and common envelope (CE) \cite{Hurley_2002}.

In a binary system, each star is surrounded by a region of space known as its Roche lobe,  which defines the region within which material is gravitationally bound to the star. When one of the stars expands and fills its Roche lobe, the material  flows to its companion  in a process known as RLOF,  which can significantly alter the mass, structure, and future evolution of both stars.  In some cases, the star can expand  beyond its Roche lobe and engulf its companion within a shared gaseous envelope. This stage of evolution is called CE. The friction between the two cores (one of each star) and the envelope causes the  binary orbit to shrink, while some of the orbital energy is transferred to the envelope, which may eventually be ejected. If the envelope is not ejected, the binary system merges prematurely
giving birth to a single BH. In contrast, if the envelope is successfully ejected,
the CE phase leaves behind a binary with final separation of just few solar radii, much smaller than the initial
one \cite{Ivanova_2013}. This can lead to the formation of close compact object binaries that eventually merge and produce GWs. 

 However, this seemingly straightforward framework is rich in uncertainties. In particular, the CE phase remains one of the most uncertain stages of binary evolution.

 The CE phase is commonly modeled using the so-called \textit{alpha prescription}, based on the conservation of energy \cite{Politano_2021}. In this prescription, given a primary star of mass $m_\mathrm{p}$ with a core of mass $m_\mathrm{c}$ and a secondary star of mass $m_\mathrm{s}$ orbiting with an initial separation $a_i$, the can eject an envelope with gravitational binding energy $E_\mathrm{bin}$ end their CE phase with an orbital separation $a_f$ if
\begin{equation}
    \alpha_\mathrm{CE} \tonde{\frac{G m_\mathrm{p} m_\mathrm{s}}{2a_i} - \frac{G m_\mathrm{c} m_\mathrm{s}}{2a_f}} = E_\mathrm{bin} .
\end{equation}

 Hence $\alpha_\mathrm{CE}$, called the CE efficiency parameter, quantifies the amount of orbital energy released during the inspiral which is successfully absorbed by the envelope.

 Unfortunately, numerical simulations \cite{Fragos_2019, Iorio_2023} showed that the production of binaries analogous to those detected by LVK often requires 
\begin{equation}
    \alpha_\mathrm{CE} >1 ,
\end{equation}
 meaning that the envelope absorbs an energy amount greater than that released by the orbital inspiral of the binary.

 This is clearly an unphysical statement, and it indicates that some energy sources (such as thermal energy) are not accounted for in the current CE formalism \cite{De_Marco_2011} or that, most likely, the \textit{alpha prescription} is not an appropriate formalism to describe such a complex phase in binary evolution.

\subsection{ The zoology of dynamical  assembly}
In a dense stellar environment, a BH experiences  numerous close interactions with other objects, leading to a  variety of dynamical processes.

A close interaction between two objects has two main effects: the system loses energy through gravitational waves production, and tidal interactions become important \cite{Samsing_2017}. Both of these processes can favor the formation of binaries.

 Additionally, a formed binary will interact frequently with single bodies , with different outcomes depending on the masses and energies of the objects at play:
\begin{enumerate}[i)]
    \item  Ionization: The binary can be widened or destroyed  by the three-body interaction. The three objects can therefore leave the interaction as single objects.
    \item  Hardening: The binary can become even more bound by transferring some of its energy to the third body. 
    \item  Exchange: One of the members of the binary can be replaced by the intruder. For example, when a binary composed of a BH and a low-mass star undergoes an encounter with a single BH, this can lead to the formation of a BBH \cite{Ziosi_2014}.
\end{enumerate}

 Repeated interactions of a tightly bound binary with third bodies cause the binary to harden further. Eventually, the system becomes so tight that its evolution is governed by gravitational wave emission, leading to its merger.

This implies that a merging BBH may not have formed in the same binary at all, and the two BHs are, instead, remnants of
independent massive stars (maybe single, or each evolved in its own separate binary) which dynamically find each other later during their lifetime.

Moreover, if the environment is dense enough, the merger product  may undergo further binary formation, hardening and coalescence. This can lead to a chain of events where we dynamically form more and more massive BBHs which produce more and more massive remnants. This process is called hierarchical merging.

The  process of dynamical assembly is facilitated in the densest stellar environments in the universe: star clusters. There are different kinds of clusters which are classified based on their mass and age. 
Globular clusters (GC) are the oldest clusters, with typical ages of $\SI{12}{\giga\year}$ and typical masses of $10^4-10^6\Msun$. They are quite compact and they typically found in galactic haloes. 
Young clusters (YSC) have formed recently (they have ages of about $\SI{100}{\mega\year}$) and they are gravitationally loose. They have typical masses of $10^2-10^5 \Msun$.
Open clusters (OSC) are similar to YSC but they are smaller and older: they have typical masses of $10^1-10^4 \Msun$ and typical ages of a few $\si{\giga \year}$.
Nuclear clusters (NSC) are found at the center of most galaxies. They have typical masses $10^4-10^8\Msun$ and typical ages of a few $\si{\giga \year}$, and they can sometimes coexist with a SMBH in a phase of mass accretion, which is called active galactic nucleus (AGN).

Estimating the effect of dynamics of the population of observable BBH mergers is no easy task. The gravitational three-body problem is notoriously chaotic \cite{Musielak_2014}, meaning that the orbits of three bodies that are under the influence of each other's gravitational potential are not predictable analytically. By extension, constructing analytical descriptions for an $N$-body system such as a stellar cluster is unfeasible\footnote{A convergent power series solution for arbitrarily many mass points that attract each according to
 Newton’s law was found in \cite{Wang_1991}, but the solution is too difficult to implement and slow to converge.}. 

Direct astronomical observations of real star clusters, on the other hand, are not fully informative on dynamics as the phenomena we are interested in happen on timescales much longer than human lifetimes. For instance, it's possible to detect stellar binaries \cite{Sana_2012} and triples \cite{Rappaport_2024}, but it's not possible to witness their history nor their future evolution. In other words, any astronomical observation is simply a snap-shot of a certain system and it does not provide direct evidence about its dynamics.

It follows that the most versatile approach to tackle the problem is by means of computational  post-Newtonian simulations.

\section{Conclusions: The Newtonian framework in a\\ post-Newtonian world}

Thanks to GW observations, we are now more knowledgeable than ever before in the field of BH astrophysics. The groundbreaking detections by the LVK collaboration have opened a new window onto the Universe, confirming the existence of BBH systems and providing unprecedented insights into their properties. 

Nonetheless, some questions still remain unanswered. In particular, there is no unequivocal interpretation for BBH mergers in which one or both components have mass in the pair-instability mass gap (approximately $\quadre{60, 120} \Msun$). 

The two primary channels proposed for BBH formation are isolated binary evolution and dynamical assembly. The isolated evolution channel faces significant challenges in explaining the observed mergers, requiring complex and often poorly understood processes like CE evolution. Even then, current models often require unphysical parameters (e.g., a CE efficiency greater than 1) to produce binaries tight enough to merge via GW emission.

Dynamical assembly, on the other hand, offers a compelling alternative. In dense stellar environments like star clusters, frequent gravitational interactions can lead to binary formation, hardening, and exchange events. These processes can ultimately drive binaries to coalesce, even potentially forming hierarchical mergers where the remnant of one merger participates in subsequent mergers, building up to higher BH masses.

For this purpose, it is important to simulate dynamics in stellar clusters, for which classical Newtonian gravity is an appropriate and useful paradigm. Although GWs are intrinsically a relativistic phenomenon, the vast majority of interactions within a star cluster occur in the weak-field, slow-motion regime. In this context, the equations of motion for an $N$-body system are adequately described by Newton's law of gravity. 

The use of Newtonian gravity significantly simplifies the computational task. Solving the full GR equations for a system of millions of stars is computationally prohibitive. Instead, specialized $N$-body codes rely on Newtonian dynamics to model the long-term evolution of star clusters.

While the Newtonian approximation is highly effective for weak-field and slow-motion systems, it does have limitations. For example, the very close encounters and the final stages of BBH inspiral and merger, do require post-Newtonian corrections (up to 3.5PN order for accurate waveform modeling). 

However, the overall dynamical evolution of the cluster, and the processes leading to the formation of the binary in the first place, are predominantly governed by Newtonian gravity. The combination of Newtonian $N$-body simulations to capture the large-scale cluster dynamics, coupled with post-Newtonian treatments for the final inspiral phase, provides a powerful framework for understanding the origin of the BBH mergers observed by LVK.  
Future observations from space-based detectors like LISA will extend the gravitational wave spectrum to lower frequencies, probing binaries in even more massive and distant environments. Incorporating Newtonian $N$-body simulations with post-Newtonian corrections will be essential for interpreting these signals and refining models of hierarchical black hole formation.

\section*{Funding}
The author acknowledges financial support from the European Research Council for the ERC Consolidator grant DEMOBLACK, under contract no. 770017, and from the German Excellence Strategy via the Heidelberg Cluster of Excellence (EXC 2181‑390900948) STRUCTURES.

\section*{Acknowledgments}
The author thanks the anonymous referees for their constructive and insightful comments that improved this manuscript. The author thanks Paolo Bussotti for their useful input.

\bibliographystyle{unsrt}
\bibliography{bibliography}

\begin{thebibliography}{10}

\bibitem{Principia}
Isaac {Newton}.
\newblock {\em {The Principia : mathematical principles of natural philosophy}}.
\newblock 1999.

\bibitem{GR_1915}
Albert {Einstein}.
\newblock {Zur allgemeinen Relativit{\"a}tstheorie}.
\newblock {\em Sitzungsberichte der K\&ouml;niglich Preussischen Akademie der Wissenschaften}, pages 778--786, January 1915.

\bibitem{John_Michell}
{Michell, J}.
\newblock On the means of discovering the distance, magnitude, \&c. of the fixed stars.
\newblock {\em Philosophical Transactions of the Royal Society}, 74:35--54, 1784.
\newblock {In a letter to Henry Cavendish}.

\bibitem{article_Mitchell}
{Chodos, A.}
\newblock {November 27, 1783: John Michell anticipates black holes}.
\newblock {\em American Physical Society News}, 18(10), 2009.

\bibitem{Genzel_2021_SMBH}
{Genzel, R.}
\newblock {\em Massive Black Holes: Evidence, Demographics and Cosmic Evolution}, pages 93--119.
\newblock Springer International Publishing, Cham, 2021.

\bibitem{Woosley_2002}
S.~E. Woosley, A.~Heger, and T.~A. Weaver.
\newblock The evolution and explosion of massive stars.
\newblock {\em Rev. Mod. Phys.}, 74:1015--1071, Nov 2002.

\bibitem{Spera_2017}
Mario {Spera} and Michela {Mapelli}.
\newblock {Very massive stars, pair-instability supernovae and intermediate-mass black holes with the sevn code}.
\newblock {\em Monthly Notices of the Royal Astronomical Society,}, 470(4):4739--4749, October 2017.

\bibitem{PISN}
{Hirschi, R.}
\newblock {Very Massive and Supermassive Stars: Evolution and Fate}.
\newblock In {\em Handbook of Supernovae}, page 567. Springer International Publishing, 2017.

\bibitem{Costa_et_al_2020}
Guglielmo {Costa}, Alessandro {Bressan}, Michela {Mapelli}, Paola {Marigo}, Giuliano {Iorio}, and Mario {Spera}.
\newblock {Formation of GW190521 from stellar evolution: the impact of the hydrogen-rich envelope, dredge-up, and $^{12}$C({\ensuremath{\alpha}}, {\ensuremath{\gamma}})$^{16}$O rate on the pair-instability black hole mass gap}.
\newblock {\em Monthly Notices of the Royal Astronomical Society}, 501(3):4514--4533, March 2021.

\bibitem{Farmer_2019}
R.~Farmer, M.~Renzo, S.~E. de~Mink, P.~Marchant, and S.~Justham.
\newblock Mind the gap: The location of the lower edge of the pair-instability supernova black hole mass gap.
\newblock {\em The Astrophysical Journal}, 887(1):53, dec 2019.

\bibitem{Hartle_2013}
{Hartle, J. B.}
\newblock {\em Gravity: An Introduction to Einstein's General Relativity: Pearson New International Edition}.
\newblock Pearson Education Limited, 2013.

\bibitem{roadmapGW_Bailes}
{Bailes, M.}, {Berger, B. K.}, and {Brady, P. R.} et~al.
\newblock Gravitational-wave physics and astronomy in the 2020s and 2030s.
\newblock {\em {Nature Reviews Physics}}, 3:344–366, 2021.

\bibitem{maggiore1_GW}
Michele {Maggiore}.
\newblock {\em Gravitational Waves: Volume 1: Theory and Experiments}.
\newblock Gravitational Waves. Oxford University Press, 2008.

\bibitem{proton_radius_2022}
{Lin, Y.-H.}, {Hammer, H.-W.}, and {Mei\ss{}ner, U.-G.}S.
\newblock New insights into the nucleon's electromagnetic structure.
\newblock {\em Physical Review Letter}, 128:052002, Feb 2022.

\bibitem{GWTC3_first}
R.~Abbott, {LIGO Scientific Collaboration}, {Virgo Collaboration}, and {KAGRA Collaboration}.
\newblock Gwtc-3: Compact binary coalescences observed by ligo and virgo during the second part of the third observing run.
\newblock {\em Physical Review X}, 13:041039, Dec 2023.

\bibitem{MANDEL20221}
Ilya Mandel and Alison Farmer.
\newblock Merging stellar-mass binary black holes.
\newblock {\em Physics Reports}, 955:1--24, 2022.
\newblock Merging Stellar-Mass binary black holes.

\bibitem{GWTC-1}
{Abbott, R.}, {LIGO Scientific Collaboration}, and {Virgo Collaboration}.
\newblock {Binary Black Hole Mergers in the First Advanced LIGO Observing Run}.
\newblock {\em Physical Review X}, 6(4):041015, October 2016.

\bibitem{GWTC3_second}
R.~Abbott, {LIGO Scientific Collaboration}, {Virgo Collaboration}, and {KAGRA Collaboration}.
\newblock Population of merging compact binaries inferred using gravitational waves through gwtc-3.
\newblock {\em Physical Review X}, 13:011048, Mar 2023.

\bibitem{Antonini_2020}
Fabio {Antonini} and Mark {Gieles}.
\newblock {Merger rate of black hole binaries from globular clusters: Theoretical error bars and comparison to gravitational wave data from GWTC-2}.
\newblock {\em Physical Review D}, 102(12):123016, December 2020.

\bibitem{Antonini_2023}
Fabio {Antonini}, Mark {Gieles}, Fani {Dosopoulou}, and Debatri {Chattopadhyay}.
\newblock {Coalescing black hole binaries from globular clusters: mass distributions and comparison to gravitational wave data from GWTC-3}.
\newblock {\em Monthly Notices of the Royal Astronomical Society}, 522(1):466--476, June 2023.

\bibitem{Antonini_2016}
Fabio Antonini and Frederic~A. Rasio.
\newblock Merging black hole binaries in galactic nuclei: Implications for advanced-ligo detections.
\newblock {\em The Astrophysical Journal}, 831(2):187, November 2016.

\bibitem{Arcasedda_2023}
Manuel {Arca Sedda}, Michela {Mapelli}, Matthew {Benacquista}, and Mario {Spera}.
\newblock {Isolated and dynamical black hole mergers with B-POP: the role of star formation and dynamics, star cluster evolution, natal kicks, mass and spins, and hierarchical mergers}.
\newblock {\em Monthly Notices of the Royal Astronomical Society}, 520(4):5259--5282, April 2023.

\bibitem{Kritos_2023}
Konstantinos Kritos, Emanuele Berti, and Joseph Silk.
\newblock Massive black hole assembly in nuclear star clusters.
\newblock {\em Physical Review D}, 108:083012, Oct 2023.

\bibitem{Fragione_2023}
Giacomo {Fragione} and Frederic~A. {Rasio}.
\newblock {Demographics of Hierarchical Black Hole Mergers in Dense Star Clusters}.
\newblock {\em The Astrophysical Journal}, 951(2):129, July 2023.

\bibitem{fastcluster2021}
Michela Mapelli, Marco Dall’Amico, Y.~Bouffanais, Nicola Giacobbo, Manuel Arca Sedda, Maria~Celeste Artale, A.~{Ballone}, U.~N. {Di Carlo}, G.~{Iorio}, F.~{Santoliquido}, and S.~{Torniamenti}.
\newblock Hierarchical black hole mergers in young, globular and nuclear star clusters: the effect of metallicity, spin and cluster properties.
\newblock {\em Monthly Notices of the Royal Astronomical Society}, 505(1):339–358, May 2021.

\bibitem{fastcluster2022}
Michela Mapelli, Y.~{Bouffanais}, F.~{Santoliquido}, M.~{Arca Sedda}, and M.~C. {Artale}.
\newblock {The cosmic evolution of binary black holes in young, globular, and nuclear star clusters: rates, masses, spins, and mixing fractions}.
\newblock {\em Monthly Notices of the Royal Astronomical Society}, 511(4):5797--5816, 02 2022.

\bibitem{Vaccaro_2023}
Maria~Paola {Vaccaro}, Michela {Mapelli}, Carole {Périgois}, Dario {Barone}, Maria~Celeste {Artale}, Marco {Dall’Amico}, Giuliano {Iorio}, and Stefano {Torniamenti}.
\newblock Impact of gas hardening on the population properties of hierarchical black hole mergers in active galactic nucleus disks.
\newblock {\em Astronomy \& Astrophysics}, 685:A51, 2024.

\bibitem{Wang_2015}
Long Wang, Rainer Spurzem, Sverre Aarseth, Keigo Nitadori, Peter Berczik, M.~B.~N. Kouwenhoven, and Thorsten Naab.
\newblock nbody6++gpu: ready for the gravitational million-body problem.
\newblock {\em Monthly Notices of the Royal Astronomical Society}, 450(4):4070–4080, May 2015.

\bibitem{Blanchet_2014}
Luc {Blanchet}.
\newblock {Gravitational Radiation from Post-Newtonian Sources and Inspiralling Compact Binaries}.
\newblock {\em Living Reviews in Relativity}, 17(1):2, December 2014.

\bibitem{Rindler_2010}
Wolfgang Rindler.
\newblock {\em Relativity: Special, General, And Cosmological Second Edition}.
\newblock Oxford University Press, 04 2006.

\bibitem{Trani_2023}
Alessandro~A. {Trani} and Mario {Spera}.
\newblock {Modeling gravitational few-body problems with tsunami and okinami}.
\newblock In Dmitry {Bisikalo}, Dmitri {Wiebe}, and Christian {Boily}, editors, {\em The Predictive Power of Computational Astrophysics as a Discover Tool}, volume 362 of {\em IAU Symposium}, pages 404--409, January 2023.

\bibitem{Hurley_2000}
Jarrod~R. Hurley, Onno~R. Pols, and Christopher~A. Tout.
\newblock {Comprehensive analytic formulae for stellar evolution as a function of mass and metallicity}.
\newblock {\em Monthly Notices of the Royal Astronomical Society}, 315(3):543--569, 07 2000.

\bibitem{Hurley_2002}
Jarrod~R. Hurley, Christopher~A. Tout, and Onno~R. Pols.
\newblock {Evolution of binary stars and the effect of tides on binary populations}.
\newblock {\em Monthly Notices of the Royal Astronomical Society}, 329(4):897--928, 02 2002.

\bibitem{Ivanova_2013}
N~Ivanova, S~Justham, X~Chen, O~De~Marco, C~L Fryer, E~Gaburov, H~Ge, E~Glebbeek, Z~Han, X-D Li, G~Lu, T~Marsh, P~Podsiadlowski, A~Potter, N~Soker, R~Taam, T~M Tauris, E~P~J van~den Heuvel, and R~F Webbink.
\newblock Common envelope evolution: where we stand and how we can move forward.
\newblock {\em The Astronomy and Astrophysics Review}, 21(1):59, February 2013.

\bibitem{Politano_2021}
M.~{Politano}.
\newblock {The final orbital separation in common envelope evolution}.
\newblock {\em Astronomy \& Astrophysics}, 648:L6, April 2021.

\bibitem{Fragos_2019}
Tassos Fragos, Jeff~J. Andrews, Enrico Ramirez-Ruiz, Georges Meynet, Vicky Kalogera, Ronald~E. Taam, and Andreas Zezas.
\newblock The complete evolution of a neutron-star binary through a common envelope phase using 1d hydrodynamic simulations.
\newblock {\em The Astrophysical Journal Letters}, 883(2):L45, oct 2019.

\bibitem{Iorio_2023}
Giuliano {Iorio}, Michela {Mapelli}, Guglielmo {Costa}, Mario {Spera}, Gast{\'o}n~J. {Escobar}, Cecilia {Sgalletta}, Alessandro~A. {Trani}, Erika {Korb}, Filippo {Santoliquido}, Marco {Dall'Amico}, Nicola {Gaspari}, and Alessandro {Bressan}.
\newblock {Compact object mergers: exploring uncertainties from stellar and binary evolution with SEVN}.
\newblock {\em Monthly Notices of the Royal Astronomical Society}, 524(1):426--470, September 2023.

\bibitem{De_Marco_2011}
Orsola De~Marco, Jean-Claude Passy, Maxwell Moe, Falk Herwig, Mordecai-Mark Mac~Low, and Bill Paxton.
\newblock On the alpha formalism for the common envelope interaction: The alpha formalism.
\newblock {\em Monthly Notices of the Royal Astronomical Society}, 411(4):2277–2292, January 2011.

\bibitem{Samsing_2017}
{Samsing, J.}, {MacLeod, M.}, and {Ramirez-Ruiz, E.}
\newblock Formation of tidal captures and gravitational wave inspirals in binary-single interactions.
\newblock {\em The Astrophysical Journal}, 846(1):36, aug 2017.

\bibitem{Ziosi_2014}
Brunetto~Marco Ziosi, Michela Mapelli, Marica Branchesi, and Giuseppe Tormen.
\newblock {Dynamics of stellar black holes in young star clusters with different metallicities – II. Black hole–black hole binaries}.
\newblock {\em Monthly Notices of the Royal Astronomical Society}, 441(4):3703--3717, 06 2014.

\bibitem{Musielak_2014}
Z~E Musielak and B~Quarles.
\newblock The three-body problem.
\newblock {\em Reports on Progress in Physics}, 77(6):065901, June 2014.

\bibitem{Wang_1991}
Qiu-Dong {Wang}.
\newblock {The global solution of the n-body problem}.
\newblock {\em Celestial Mechanics and Dynamical Astronomy}, 50(1):73--88, January 1991.

\bibitem{Sana_2012}
H.~{Sana}, S.~E. {de Mink}, A.~{de Koter}, N.~{Langer}, C.~J. {Evans}, M.~{Gieles}, E.~{Gosset}, R.~G. {Izzard}, J.~B. {Le Bouquin}, and F.~R.~N. {Schneider}.
\newblock {Binary Interaction Dominates the Evolution of Massive Stars}.
\newblock {\em Science}, 337(6093):444, July 2012.

\bibitem{Rappaport_2024}
{Rappaport, S. A.}, {Borkovits, T.}, {Mitnyan, T.}, {Gagliano, R.}, {Eisner, N.}, {Jacobs, T.}, {Tokovinin, A.}, {Powell, B.}, {Kostov, V.}, {Omohundro, M.}, {Kristiansen, M. H.}, {Jayaraman, R.}, {Terentev, I.}, {Schwengeler, H. M.}, {LaCourse, D.}, {Garai, Z.}, {Pribulla, T.}, {Maxted, P. F. L.}, {Bíró, I. B.}, {Csányi, I.}, {Pál, A.}, and {Vanderburg, A.}
\newblock Seven new triply eclipsing triple star systems.
\newblock {\em Astronomy \& Astrophysics}, 686:A27, 2024.

\end{thebibliography}

\end{document}